\documentclass[preprint]{aastex}

\usepackage{color}
\usepackage{appendix}
\usepackage{natbib}
\usepackage{rotating} 
\usepackage{lscape}
\usepackage{graphicx}

\begin{document}
\title{Properties of M31. V: 298 Eclipsing Binaries from PAndromeda}

\shorttitle{M31 Eclipsing binaries} 

\author{C.-H. Lee\altaffilmark{1,2}, J. Koppenhoefer\altaffilmark{2,1}, S. Seitz\altaffilmark{1,2}, R. Bender\altaffilmark{1,2}, A. Riffeser\altaffilmark{1,2}, M. Kodric\altaffilmark{1,2}, U. Hopp\altaffilmark{1,2}, J. Snigula\altaffilmark{2,1}, C. G\"ossl\altaffilmark{1,2}, R.-P. Kudritzki\altaffilmark{3}, W. Burgett\altaffilmark{3}, K. Chambers\altaffilmark{3}, K. Hodapp\altaffilmark{3}, N. Kaiser\altaffilmark{3}, C. Waters\altaffilmark{3}}
\altaffiltext{1}{University Observatory Munich, Scheinerstrasse 1, 81679 Munich, Germany}
\altaffiltext{2}{Max Planck Institute for Extraterrestrial Physics, Giessenbachstrasse, 85748 Garching, Germany}
\altaffiltext{3}{Institute for Astronomy, University of Hawaii at Manoa, Honolulu, HI 96822, USA}

\begin{abstract}
The goal of this work is to conduct a photometric study of eclipsing 
binaries in M31. We apply a modified box-fitting algorithm to search for 
eclipsing binary candidates and determine their period. We classify these 
candidates into detached, semi-detached, and contact systems using the 
Fourier decomposition method. We cross-match the position of our detached 
candidates with the photometry from Local Group Survey \citep{2006AJ....131.2478M} 
and select 13 candidates brighter than 20.5 magnitude in V. The relative 
physical parameters of these detached candidates are further characterized 
with Detached Eclipsing Binary Light curve fitter (DEBiL) by \cite{2005ApJ...628..411D}. 
We will follow-up the detached eclipsing 
binaries spectroscopically and determine the distance to M31.
\end{abstract}


\section{Introduction}

Eclipsing binaries are important in two aspects. First of all, they can 
provide information on the physical parameters of the system. For example, 
their light-curves can be used to derive the inclination angle of their 
orbital plane, and the stellar radius in terms of the orbital distance. The 
spectroscopic observation can be used to infer the mass of individual stars, 
the orbital distance, and stellar temperature. These information are essential 
for the theoretical study of stellar evolution.

Secondly, eclipsing binaries can serve as distance indicators. This is possible because once we 
know the stellar radius and temperature from joint photometric and 
spectroscopic observations, 
we can derive the distance ($d$) by:
\begin{equation}
f_{\lambda}=\frac{1}{d^2}(R_1^2F_{1,\lambda}+R_2^2F_{2,\lambda}) \times 10^{-0.4A(\lambda)},
\end{equation}
where $R_{1,2}$ are the radii of the primary and secondary, $F_{1,\lambda}$ and $F_{2,\lambda}$
are the surface fluxes of the primary and secondary components, and $A(\lambda)$ is the 
total extinction:
\begin{equation}
A(\lambda)=E(B-V)[k(\lambda -V)+R_v],
\end{equation}
where $E(B-V)$ is the reddening, $k(\lambda -V)$ is the normalized extinction curve defined
as $E(\lambda -V)/E(B-V)$, and $R_v$ is the ratio of total to selective extinction in V band. 
By matching the model atmospheres to the broad band photometry, we are able to determine the 
value of $E(B-V)$ and $R_v$ and estimate the extinction, which is usually hard to measure and 
assumed $R_v$=3.1.

It is worth to 
note that according to the extent of Roche-lobe filling and the deformation 
of the stellar surface, eclipsing binaries can be classified into subclass of detached, 
semi-detached and over-contact. It has been suggested that the detached eclipsing binaries 
are more suitable for distance determination \citep[see e.g.][]{1997eds..proc..273P}, 
although some authors argue that the Roche-lobe filling semi-detached eclipsing binaries 
can put more constraints on the light-curve modeling thus have advantages 
to be standard candle as well \citep{2002ApJ...571..293W}. 

Previous studies of extra-galactic distance determination used the Large Magellanic 
Cloud (LMC) as a distance anchor. While LMC bears irregularity in its three 
dimensional shape and low metalicity, many authors suggest to use M31 as a 
stepping stone to the cosmic distance determination \citep[see e.g.,][and 
reference therein]{2001ApJ...559L.109C, 2010A&A...509A..70V}. The merits of 
M31 include its simple geometry, the stars that are potential distance 
indicators in M31 are bright enough to be resolved, a local counter-part of 
the spiral galaxies that are used to determine the extra-galactic distance 
\citep[see e.g.,][]{2001ApJ...553...47F}, and a local benchmark to calibrate 
the Tully-Fisher relation. 

In order to detect the eclipsing binaries in M31, one requires to monitor a large fraction 
of M31 with high cadence. Previous studies, e.g. the \textit{DIRECT} project 
\citep{1998AJ....115.1016K,1998AJ....115.1894S,1999AJ....117.2810S,1999AJ....118..346K,1999AJ....118.2211M,2003AJ....126..175B} observed M31 with a 11' $\times $ 11' field of 
view (FOV) in 1996 and 1997, and later expanded to a 22' $\times $ 22' FOV 
in 1998 and 1999, in a mosaic way to cover an area of 0.5 degree$^2$ 
\citep{2004ASPC..310...33M}. On the other hand, \cite{2006A&A...459..321V}
use the 33.8' $\times $ 33.8' Wide Field Camera on-board the Issac Newton 
Telescope in Spain to observe the north-eastern part of M31 from 1999 to 2003 
(in total 21 nights). The limitations of these studies is that their observations 
only cover part of M31 disk. With the $\approx$ 7 degree$^2$ FOV of 
Pan-STARRS 1 (PS1), we are able to cover the entire disk of M31 in one shot \citep[see e.g.
Fig. 1 of][]{2013ApJ...777...35L} with a time 
resolution of up to 0.5 day. The PS1 started to observe M31 since 2010. Here we 
present the results from the three year PS1 monitoring. 

This paper is organized as follows. In section \ref{sec.data} 
we present our data analysis. In section \ref{sec.bls} we describe 
our procedure to search for periodic variable sources, and invented 
a modified box-fitting method to identify eclipsing binary candidates. Classifications 
of our binary candidates are presented in section \ref{sec.class}.
We cross-match detached binaries in our sample with the Local Group 
Survey to find binaries that are ideal to follow-up; we further use the 
Detached Eclipsing Binary Light curve fitter to characterize these bright 
candidates, and extract their parameters in section \ref{sec.deb}, 
followed by a summary in section \ref{sec.sum}.  

\section{Data Analysis}
\label{sec.data}
The Pan-STARRS 1 (hereafter PS1) survey \footnote{http://pan-starrs.ifa.hawaii.edu/public/}  
uses a 1.8 meter telescope equipped 
with a giga-pixel camera (GPC) located at Haleakala in Hawaii. The camera 
consists of 60 detectors, each detector is segmented in an 8 $\times$ 8 array 
of 590 $\times$ 598 pixel cells. Each pixel has a size of 10 $\mu$m, 
corresponding to 0.258''/pixel. The FOV of each detector is 20'.95 $\times$ 
20'.74 and the FOV of the 60 detectors is approximately 7 degree$^2$. 

The PS1 survey includes dedicated observations of M31, the so-called PAndromeda project. 
PAndromeda makes use of $\sim$ 2\% of the PS1 observation time, starting from 
July until December each year, to monitor M31. In this work we present the 
result of the three-year PAndromeda data, taken between 2010 and 2012. The 
original design of PAndromeda is to search for short timescale microlensing 
events, thus we monitor M31 in two time block per night, separated by three 
to five hours, to have a time resolution of less then one day. Such 
high-cadence observation strategy is also useful for short period variables. 
The two observation blocks yield 12 observations in $r_{P1}$ 
and 6 observations in $i_{P1}$, with an exposure time of 60 seconds.
The observations in each block are combined to increase 
the S/N. In the end we have 330 epochs in $r_{P1}$ and 250 epochs in $i_{P1}$.

After the images are taken with the PS1 telescope, they are processed by the 
Image Processing Pipeline \citep[IPP,][]{2006amos.confE..50M}. The pipeline 
runs several successive steps, including bias and dark correction, masking, 
artifact removing, flat-fielding, astrometric calibration 
\citep{2008IAUS..248..553M}, and resampling to a common pixel scale to a 
sky-based image plane, so-called skycells.

Afterwards, we perform Difference Imaging Analysis 
\cite[DIA,][]{1998ApJ...503..325A} using our customized software 
MDia \citep{2013ExA....35..329K} and generate light-curves of all 
point sources. Details of our analysis can be found in 
\cite{2012AJ....143...89L} and \cite{2013AJ....145..106K}.

\section{Search for Variability}
\label{sec.bls}
In order to identify all periodic signals that can be attributed to
eclipsing binaries systems we apply a detection algorithm to all 738,755
$r_{P1}$-band light curves that have been extracted from the PAndromeda
data. Our detection procedure is based on the boxfitting algorithm
\citep{2002A&A...391..369K} which has been developed to detect
box-shaped brightness decreases such as planetary transits. 
We run the boxfitting algorithm using 100001 test periods between 0.25
days and 19.1 days which are equally distributed in 1/p. The duration
of the eclipses is limited to 0.25 phase units. 
The algorithm is known to be also useful to detect detached and
semi-detached eclipsing binary systems
\cite[e.g.][]{2012MNRAS.425..950N}, however, due to the larger ratio
between the radii of the primary and secondary star the light curves
of eclipsing binaries are generally more V-shaped than in the case of
planetary transits. In order to account for that we refit the folded
light curve for each detected period with a symmetric trapezium that
has the same full width half maximum as the best fitting box. 
The approach has been used by \cite{2013A&A...560A..92Z} already.
We define the V-shape parameter as the fraction of the time spend in
ingress and egress with respect to the total duration of the
eclipse:\\\\
\begin{equation}
V = ( T_2 - T_1 + T_4 - T_3 ) / ( T_4 - T_1 ) ,
\end{equation}
with T$_1$ and T$_2$ being the phase values at the beginning and end
of the ingress and T$_3$ and T$_4$ being the phase values at the
beginning and end of the egress. 
Note that the values of T$_1$, T$_2$, T$_3$, and T$_4$
are parameters that are fitted when applying the detection algorithm 
and they do not relate in a direct way to the realy start/end of the 
ingress/egress of the light curve since the fitted trapezium model is
only an approximation of the real brightness variation.
With this definition, $V$ = 0
corresponds to a pure box-shaped eclipse and $V$ = 1 corresponding to
an eclipse without flat central part (i.e. T$_2$ = T$_3$). In addition
we define the signal-to-noise ratio (S/N) of the detection as 
the eclipse depth $\delta$
divided by the standard deviation $\sigma_{ecl}$ of the combination of 
all points during the eclipse (i.e. all points between (
T$_1$ + T$_2$ ) / 2. and ( T$_3$ + T$_4$ ) / 2. ):\\\\
\begin{equation}           
S/N = \frac{\delta}{\sigma_{ecl}}.                           
\end{equation}            

We calculate the S/N for both the primary and the secondary
eclipse. After removing the eclipses we re-run the boxfitting 
algorithm a second time on each light curve and calculate the S/N of
the best fitting period S/N$_{removed}$. This value is a good measure
for the variation of the out-of-eclipse part of the candidate and can
be used to distinguish eclipsing binaries from other variable sources   
(see below).
        
\begin{figure}            
  \centering               
  \includegraphics[width=0.6\textwidth]{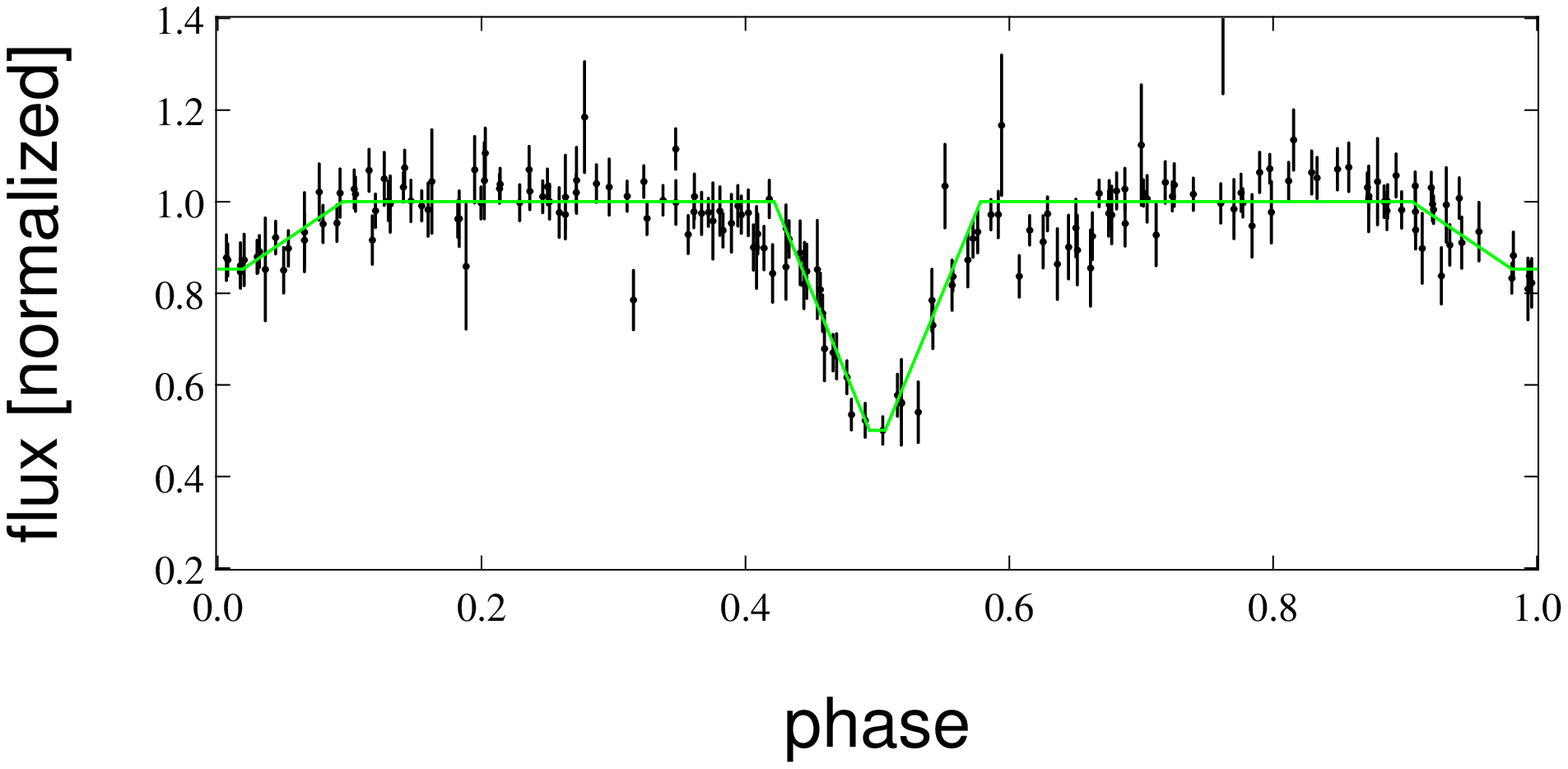} 
  \caption{Example normalized and phase-folded light curve of
one of the detached eclipsing binaries we detected in M31. The overplotted 
green line shows the best fitting trapezium model as fitted by our
detection algorithm.}               
  \label{fig.example_fit}   
\end{figure}

Eclipsing binary light curves usually show a secondary eclipse which  
is offset by 0.5 in phase units for circular orbits. We therefore fit a 
secondary eclipse for each period that has been detected by the  
boxfitting algorithm. The secondary eclipse is first fitted as a box 
which is offset by exactly 0.5 phase units from the center of the  
primary eclipse and subsequently refined to a trapezium shape in the 
same way the primary eclipse has been refined. Fig.
\ref{fig.example_fit} shows an example folded lightcurve of one of our 
detected eclipsing binaries together with the best fitting  
trapezium-shaped primary and secondary eclipse models.

Since the boxfitting
algorithm searches only for one eclipse in the folded light curve the
best fitting period for eclipsing binary light curves usually
corresponds to half the true period with the primary and secondary
transit folded together. The true binary period is found as the second
or third best fitting period. We therefore select the six best fitting
periods for each input light curve and select the period with the
lowest $\chi^2$ of the refined fit that is including a secondary
eclipse (see above).

\clearpage

We apply a number of criteria to select the best eclipsing binary
candidates. First we reject all detections with periods close to alias
periods that are enhanced by the window functions of the observations.
The alias periods are identified as peaks in the histogram of detected
periods. In addition we require a candidate to have at least 10 points
in the primary eclipse and 5 
points in the secondary eclipse. We
split the remaining detections in two samples:\\\\

\begin{itemize}
\item{The high S/N sample contains all light curves with an $RMS$ of the
    out-of-eclipse part of the light curve of lower than 0.20 and an
    effective signal-to-noise ratio S/N$_{eff}$ larger than 12.
    S/N$_{eff}$ is calculated as the sum of the S/N of primary and
    secondary eclipse reduced by S/N$_{removed}$ (see above).}
\item{The low $\chi^2$ sample contains all light curves with a reduced
    $\chi^2$ of the refined binary fit lower than 2 and a sum of the
    S/N of primary and secondary eclipse larger than 12.}\\
\end{itemize}

We visually inspected 5344 candidates in the high S/N sample and 824
candidates in the low $\chi^2$ sample and identified 298 eclipsing
binary systems. The remaining candidates were false detections due to
systematic outliers and pulsating variable stars such as
$\delta$-Cepheids. Note that we only select light curves for which
the primary and secondary eclipses show a symmetric shape
in order to have a clean sample. 

The number of 298 eclipsing binaries is a rather small
value compared with Vilardell et al. (2006). The reason for the 
relatively low number of detected eclipsing binaries is due to the 
relatively short exposure times of the current work compared to 
Vilardell et al. (2006). Note also that we only select eclipsing binary
systems with high signal to noise ratio and good light curves.

\section{Classification}
\label{sec.class}
In this section we attempt to classify our eclipsing binaries 
based on the shape of their light curves. The broadly used 
taxonomy of eclipsing systems is based on the resemblance of 
light curves to certain protypes, which are catagorized in 
Algol-type (EA), $\beta$ Lyrae-type (EB), and W Ursae-Majoris-type 
(EW), see e.g. General Catalog of Variable Stars \citep[GCVS database][]{2009yCat....102025S}. 
However, this classification scheme does 
not fully reflect the physical configuration of a binary system. 
Alternatively, we classify our binary systems based on the scheme 
of detached (ED), semi-detached (ESD), and contact (EC), which 
reflect the Roche-lobe filling status of both components.

Following the procedure of \cite{1993PASP..105.1433R} and \cite{2002AcA....52..397P}, 
we classify our candidates with the Fourier decomposition method 
by fitting the phase-folded normalized light curves with a series 
of sine and cosine functions:

\begin{equation}
f(\phi) = \Sigma_{i=1}^{4} a_i cos(2\pi i \phi) + b_i sin(2\pi i \phi)
\label{equ.fourier}
\end{equation}

where $f(\phi)$ is the normalized flux at phase $\phi$. It has been 
shown the different configurations of eclipsing binaries are separated 
on the $a_2$-$a_4$ plane. Such approach has be used to classify eclipsing
binaries in the Milky Way from the ASAS sample \citep{2002AcA....52..397P} and in the 
Magellanic Clouds from the MACHO sample \citep{2007ApJ...663..249D}. 
Before extending this classification scheme to our sample, we verify this 
classification scheme using two well-studied binaries 
in M31 \citep{2005ApJ...635L..37R,2010A&A...509A..70V}, where their 
configurations are well-known from joint analysis of 
photometric and spectroscopic analysis. The fourier decomposition method 
returns the correct configuration for these two extra-galactic binaries.
We thus adopted the boundary values of $a_2$ and 
$a_4$ from \cite{2002AcA....52..397P} and classify our candidats accordingly. 
The results can be seen in Fig. \ref{fig.a24}. To have an idea of how 
robust the $a_2$ and $a_4$ parameters are determined from the light curves,
we perform a bootstrapping test on each light curve for 100 iterations, 
and calculate the standard deviation within these 100 iterations.
The standard deviation of $a_2$ and $a_4$ is then plotted as error-bar
in Fig. \ref{fig.a24}. 

\begin{figure*}[!h]
    \centering
    \includegraphics[scale=0.6]{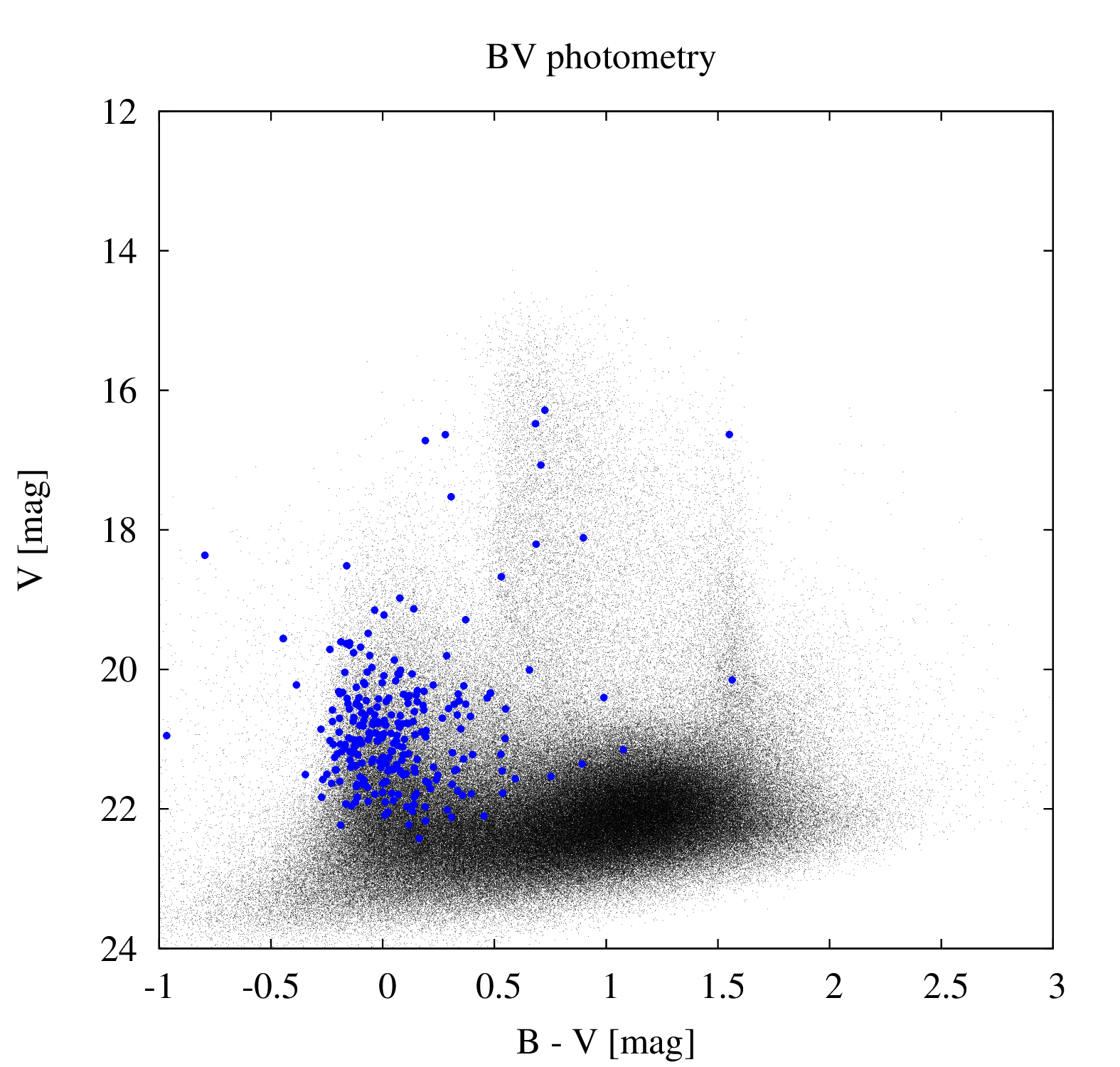}
    \caption{Color magnitude diagram for the detected eclipsing binaries. The underlaying black 
dots are photometric measurements taken from the Local Group Galaxy Survey \citep{2006AJ....131.2478M}.
The detected eclipsing binaries exhibit blue colors, allude to young, massvie components that populate
the spiral arms.}
    \label{fig.color}
\end{figure*}

\begin{figure*}[!h]
    \centering
    \includegraphics[scale=0.5]{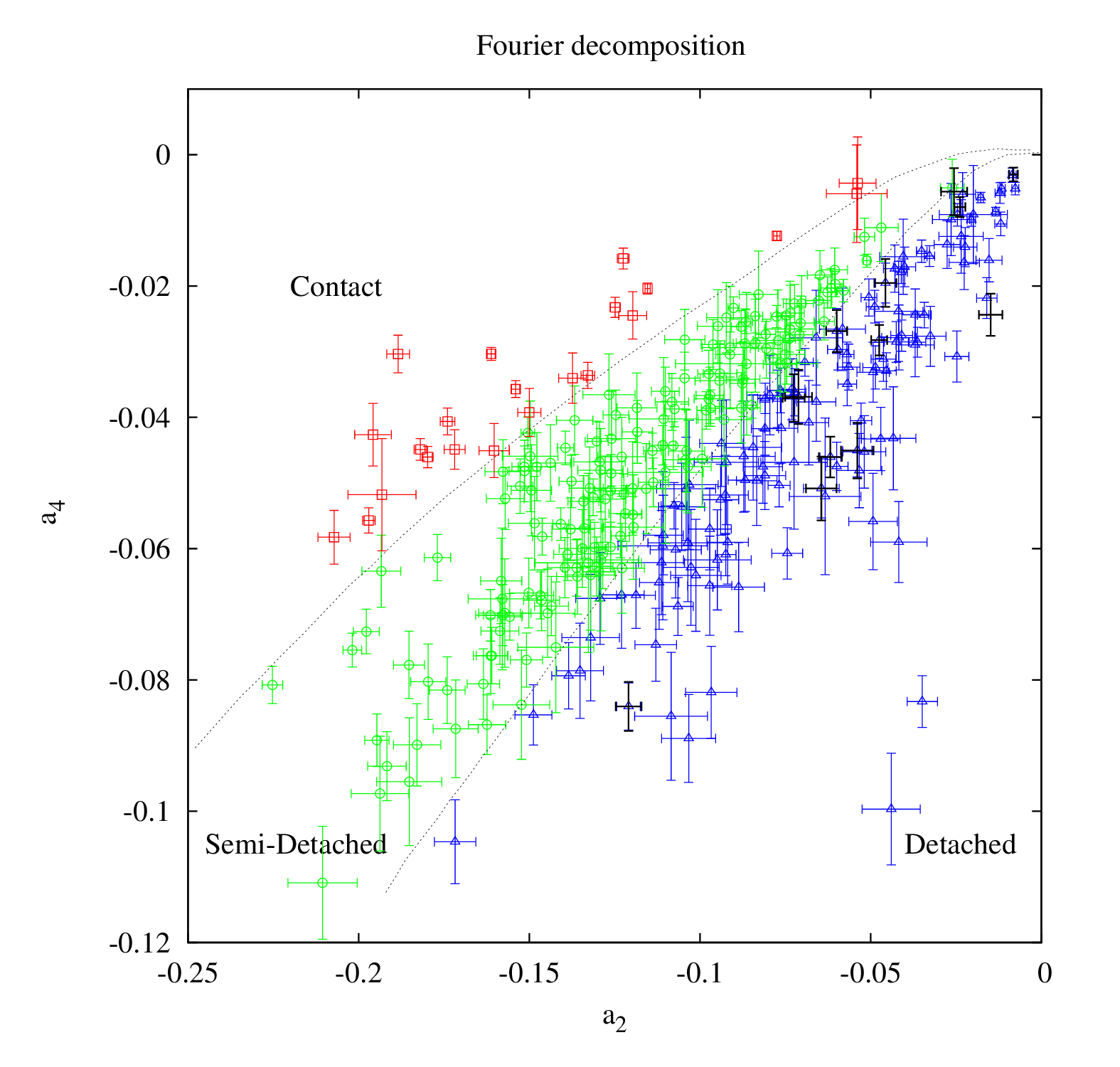}  
    \caption{Classifying binary candidates using the Fourier decomposition 
parameters $a_2$ and $a_4$ in equation \ref{equ.fourier}. Blue triangles denote 
the detached systems, green circles indicate the semi-detached systems, while 
red squares represent the contact systems. The black error bars indicate the 
bright detached systems discussed in section \ref{sec.deb}. The two dotted curves delineate the 
boundaries of different systems, as shown in \cite{2002AcA....52..397P}.}
    \label{fig.a24}  
\end{figure*}

We note that there 
are fewer contact binaries (22) than detached (120) and semi-detached (158) systems, 
in contrary to the results of Milky Way sample from the All Sky 
Automated Survey (ASAS) in \cite{2006MNRAS.368.1311P}. This might be 
contributed by the combination of intrinsic faintness of the contact 
system and the remoteness of M31. The expected 
brightness of contact binaries in M31 can be derived using the 
period-luminosity relation from \cite{1996ASPC...90..270R}:\\
\begin{equation}
M_V=-2.38\mathrm{log}P+4.26(B-V)+0.28
\end{equation}
with an 1-$\sigma$ dispersion of 0.24. Assuming a period of 1 day, 
a (B-V) color of 0.0 (see Fig. \ref{fig.color}), and given a distance modulus 
of 24.36 mag for M31 \citep{2010A&A...509A..70V}, the period-luminosity
relation predicts a observed magnitude of V=24.12 mag, which is not reached
in the PAndromeda survey. Hence, our contact binary candidates are either
more likely to reside in the foreground, or indeed belong to M31 but 
have longer periods.
In addition, our search algorithm limits the 
duration of the eclipses to be 0.25 in orbital phase; while contact binaries 
have largely sinusoidal light curves and the eclipses can span more than 
half of the orbital phase, the highly distorted contact systems can be
missed by our algorithm, hence the detection bias can affect the number of 
the contact systems as well.

\begin{figure*}[!h]
    \centering
    \includegraphics[scale=0.8]{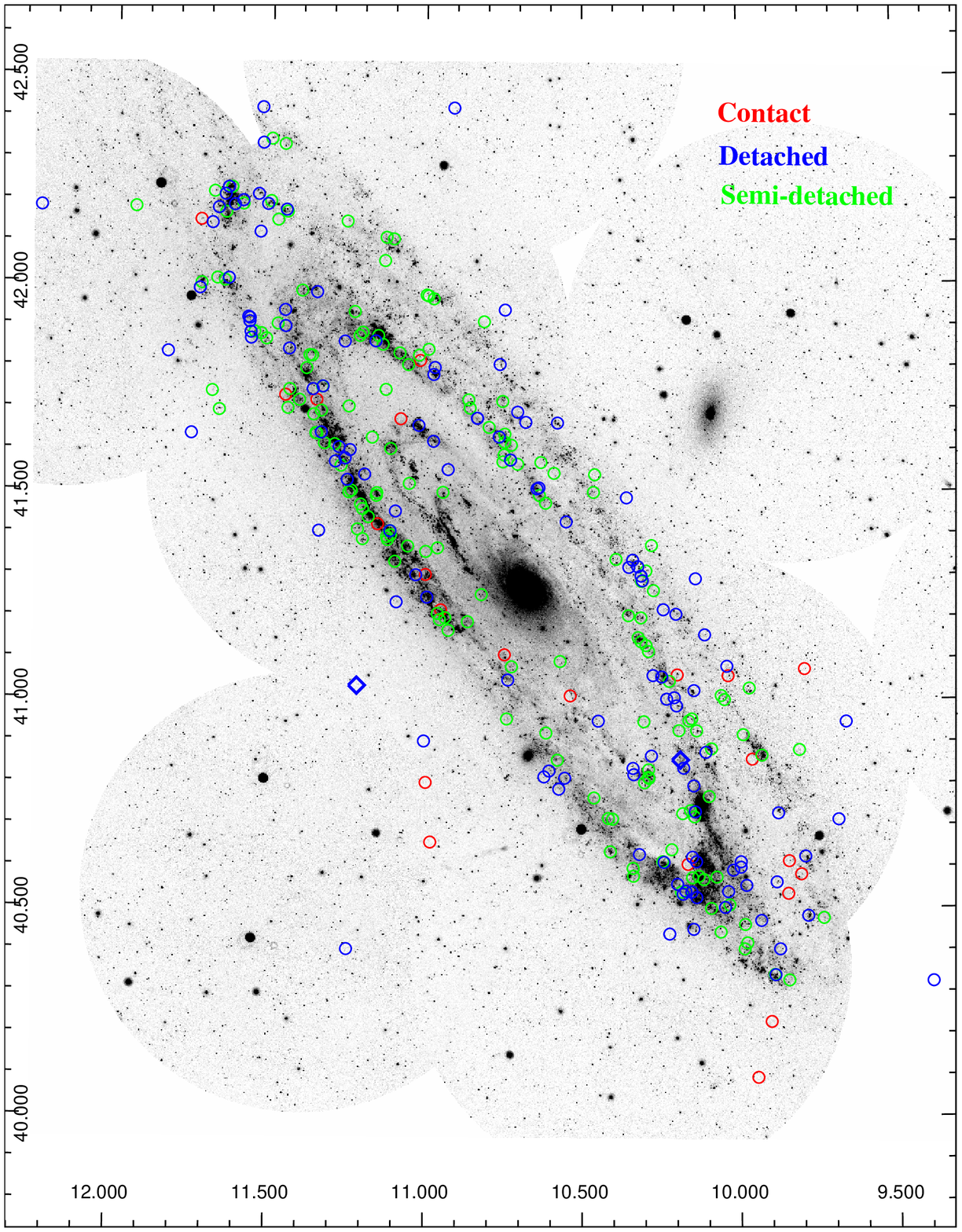}  
    \caption{Spatial distribution of different binary configurations. Contact systems are marked in red, semi-detached systems are 
marked in green, and the detached systems are marked in blue. Please note that the two foreground detached systems discussed in 
section \ref{sec.deb} are marked in rhombi. The underlying image is taken from GALEX \citep{2007ApJS..173..185G}.}
    \label{fig.spat}  
\end{figure*}

The spatial distribution of our binary candidates are shwon in Fig. \ref{fig.spat}. 
From the spatial distribution, there are some binaries locate
outside the main disk of M31, which could be attributed to foreground stars
in the Milky Way. Indeed, one of our bright detached system presented in section
\ref{sec.deb}, i.e. 036-15741, falls outside the main disk of M31, 
and is too bright/red to 
fit into the M31 scenario, hence we attribute it to be a foreground Milky Way system.
From the ASAS Milky Way sample, the detached systems are concentrated on the 
Galactic plane, suggesting that these are more massive and intrinsic 
brighter systems, while the contact systems are isotropically 
distributed \citep{2006MNRAS.368.1311P}. However, due to  our 
limited sample of contact binaries and the highly inclined disk 
of M31, we could not distinguish the difference in spatial distribution 
of these two different systems. 
In addition, we would like to note that the eclipsing binaries 
detected in the survey will likely have rather bright and blue components 
\citep[see e.g. Fig. 4 in][]{2006A&A...459..321V} and are expected to have 
young massive components, hence mostly distributed along the spiral arms. 
To verify this hypothesis, we cross-matched our eclipsing binaries with the 
photometry of \cite{2006AJ....131.2478M} to derive their colors and present 
their color-magnitude diagram in Fig. \ref{fig.color}, and find out that they 
indeed exhibit blue colors.

\begin{figure*}[!h]
    \centering
    \includegraphics[scale=0.8]{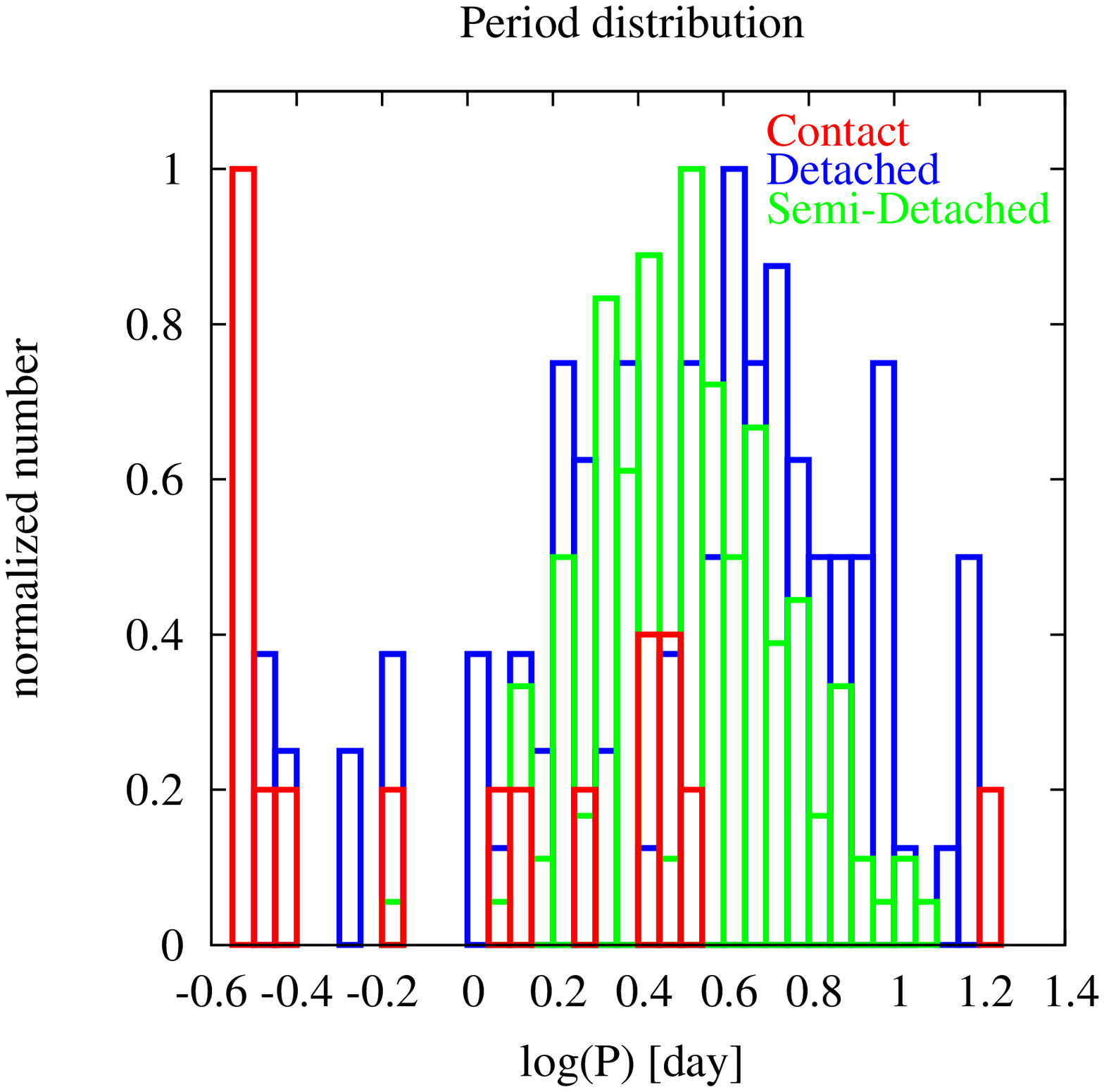}  
    \caption{Period distribution of different binary configurations. 
Contact systems are shown in red, detached systems are marked in blue, and 
the semi-detahched systems are denoted in green. Most of the contact systems 
show relatively short periods, while the detached and semi-detached systems
are peaked at period larger than 1 day.}
    \label{fig.histo_per}  
\end{figure*}

We also show the period distribution in Fig. \ref{fig.histo_per}. 
The contact systems dominate the short period range, while the detached 
and semi-detached systems are peaked at logP $>$ 1 day, in agreement 
with the results of Milky Way sample from ASAS. 
The only exception is 
a contact system with period larger than 10 days. We check the error
budget of $a_2$ and $a_4$ estimated from bootstrapping, which indicates
a solid contact classification on the $a_2$-$a_4$ plane. We note that 
there are several apparently contact systems with period larger than 
10 days reported by \cite{2001AJ....121..254R}. Such systems can be 
explained by a semi-detached scheme, where one of the component is filling
its Roche lobe and the other component is small or with lower temperature, 
thus the light curve reflects only the ellipsoidal variations of the larger, 
Roche-filling component. The color of such systems is red compared
to normal contact systems. We thus cross-match the position with the local
group survey \citep{2006AJ....131.2478M}, and found that the color of our long-period
contact binary is indeed red, with (V-I) = 0.824, and belongs to the long-period
semi-detached sheme as shown in Fig. 2 of \cite{2001AJ....121..254R}.

\clearpage

\section{Detached binaries suitable for distance determination}
\label{sec.deb}
\begin{figure*}[!h]
    \centering
    \includegraphics[scale=0.4]{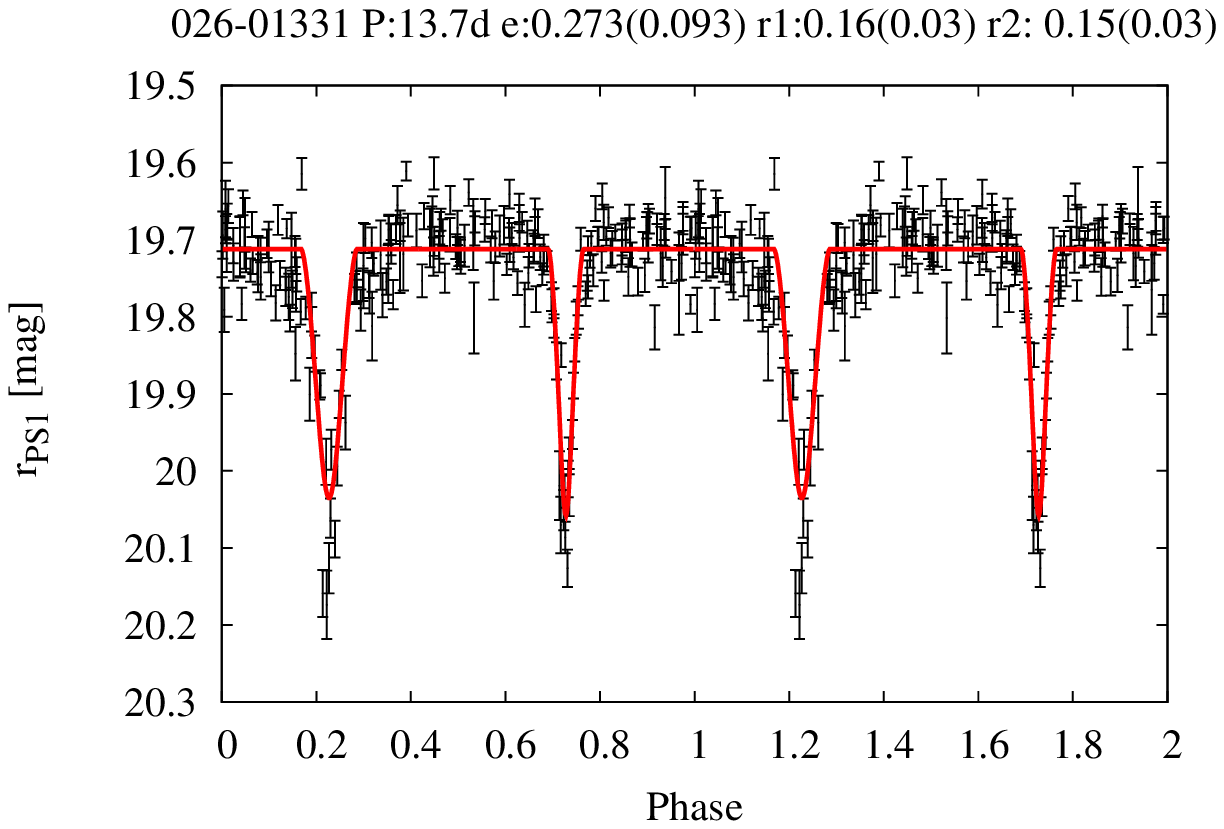}  
    \includegraphics[scale=0.4]{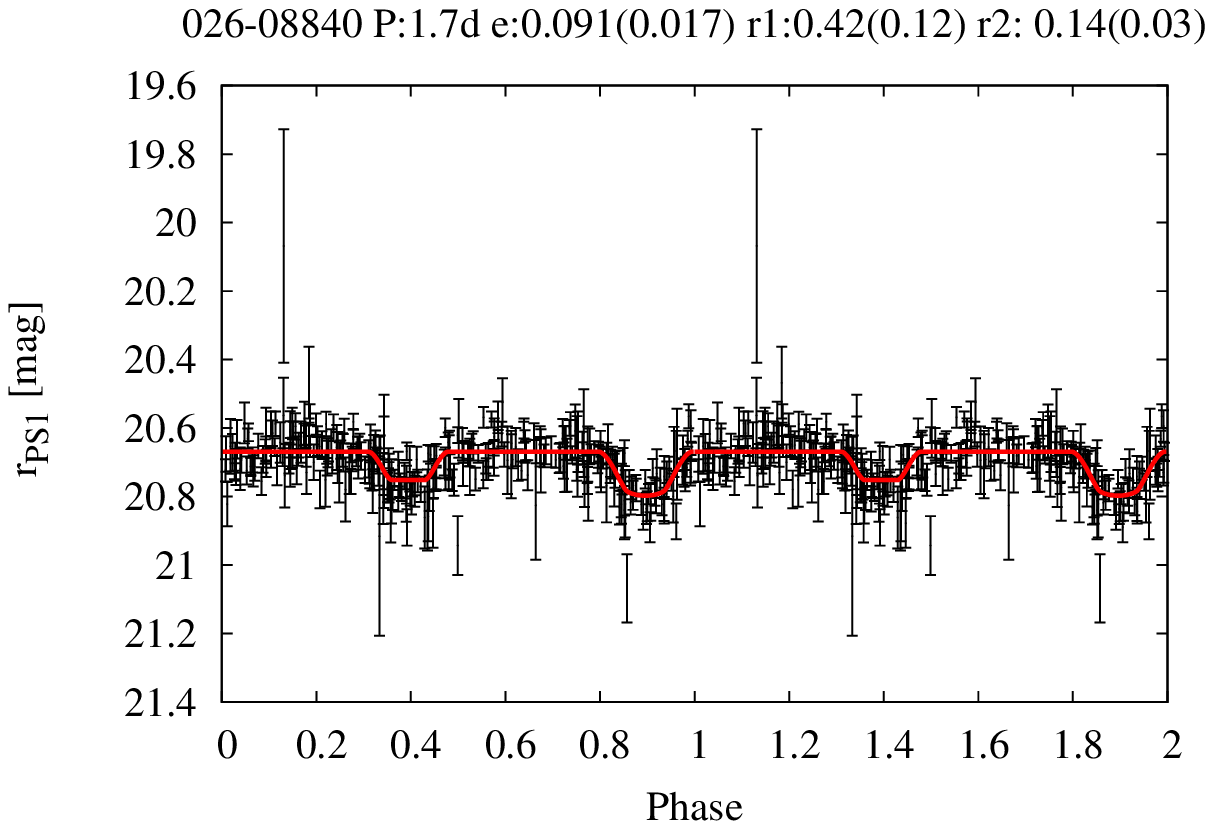}
    \includegraphics[scale=0.4]{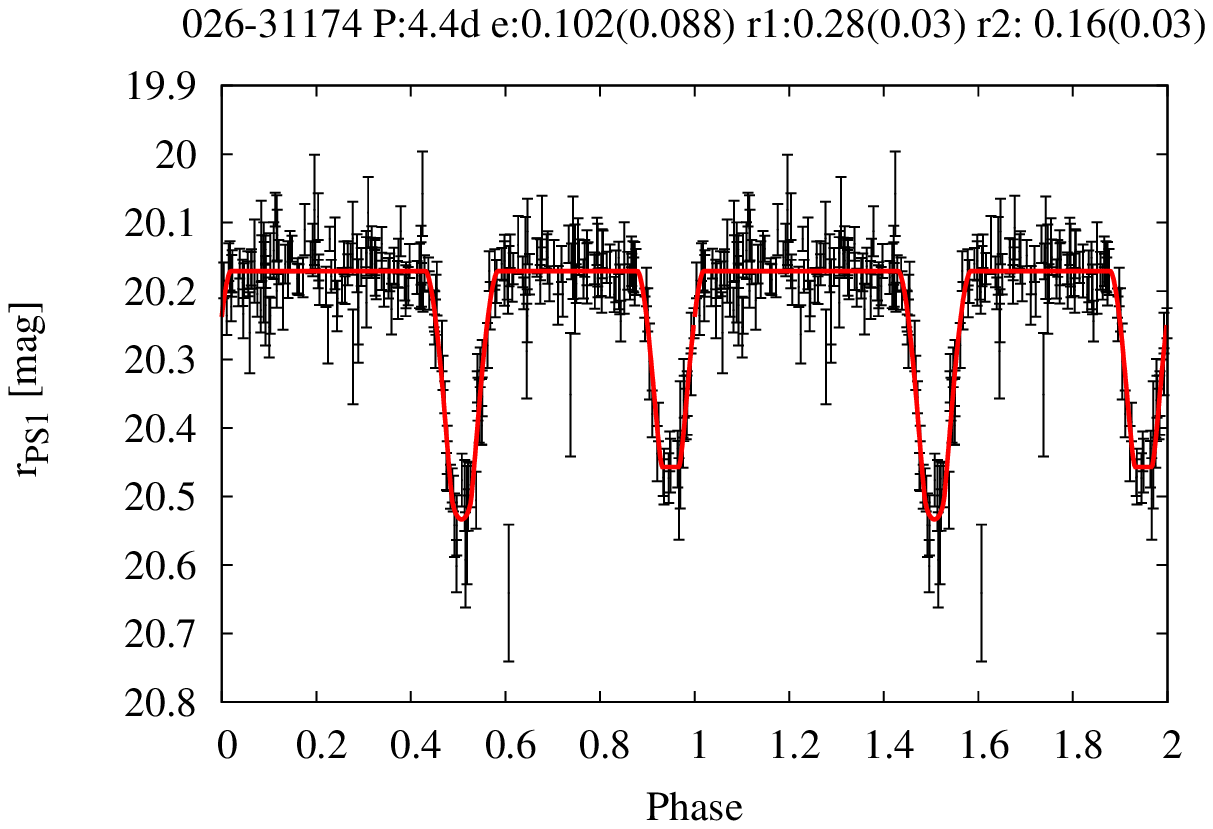}
    \includegraphics[scale=0.4]{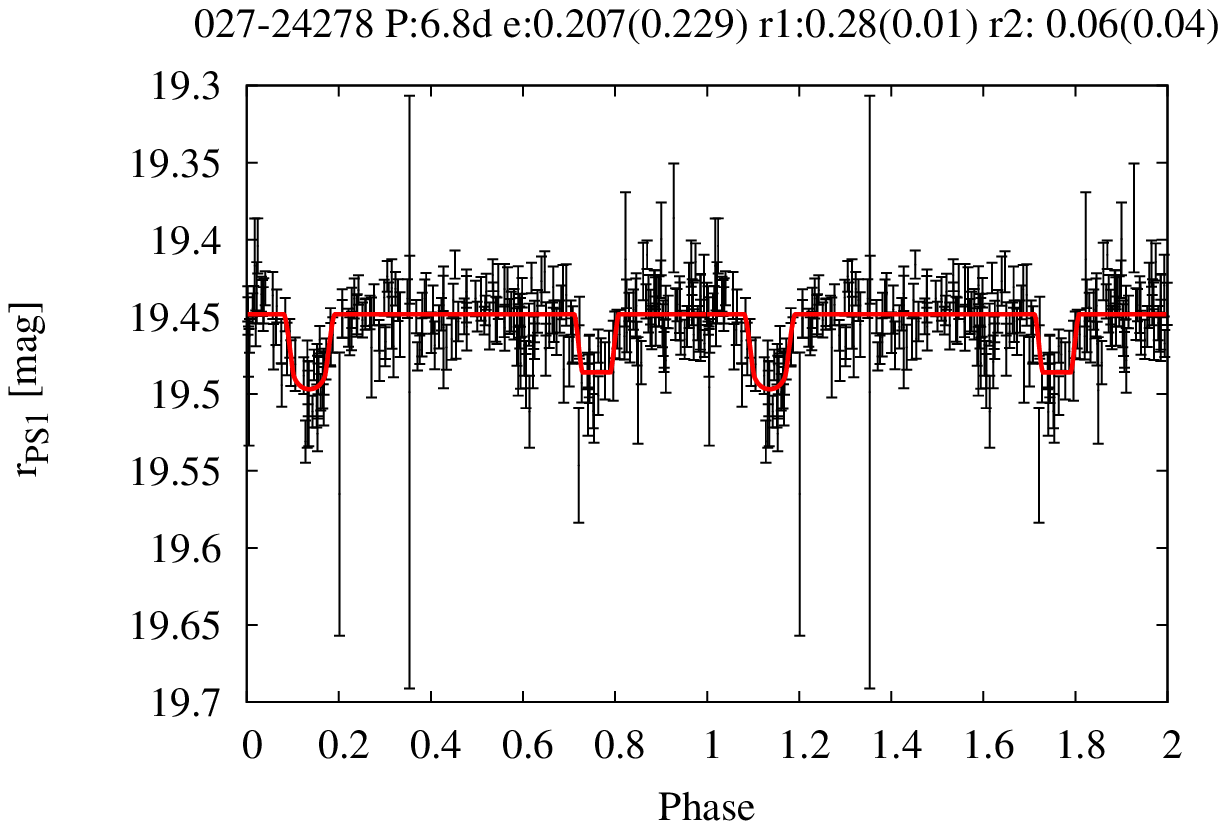}
    \includegraphics[scale=0.4]{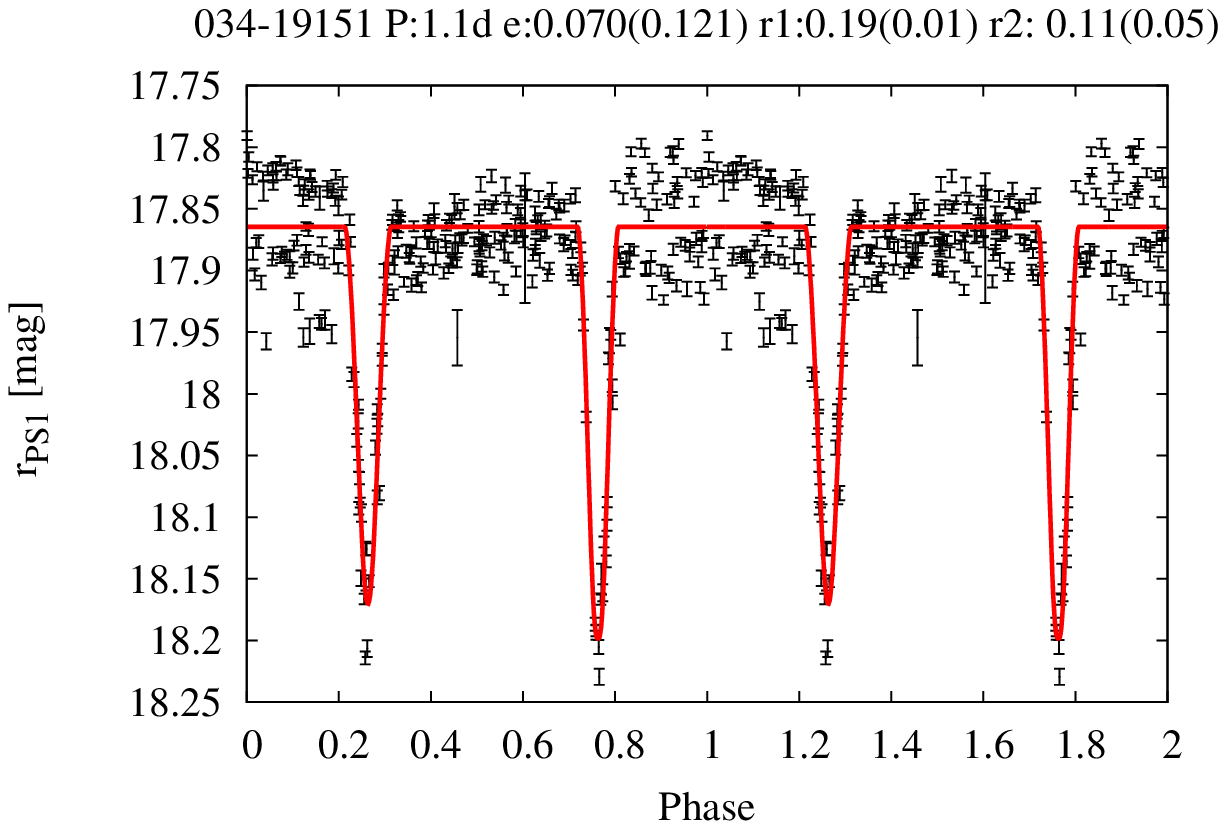}
    \includegraphics[scale=0.4]{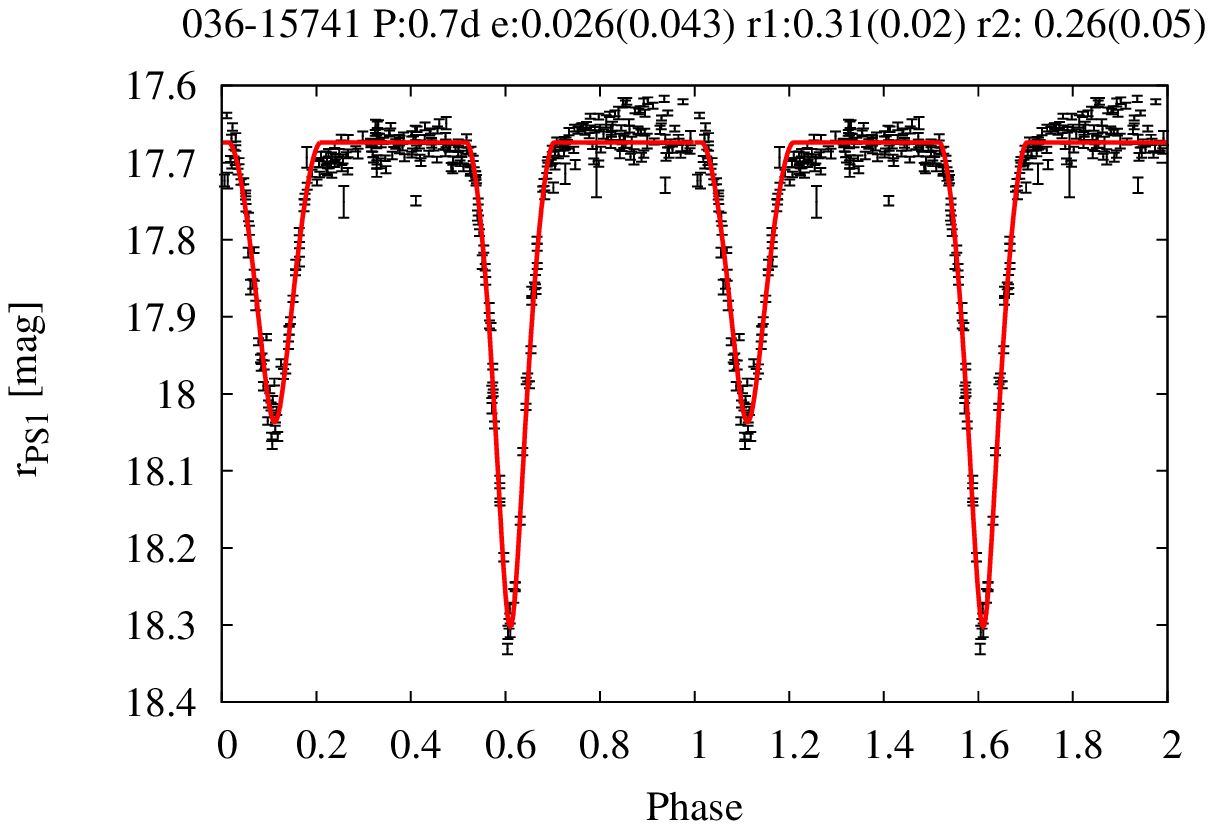}
    \includegraphics[scale=0.4]{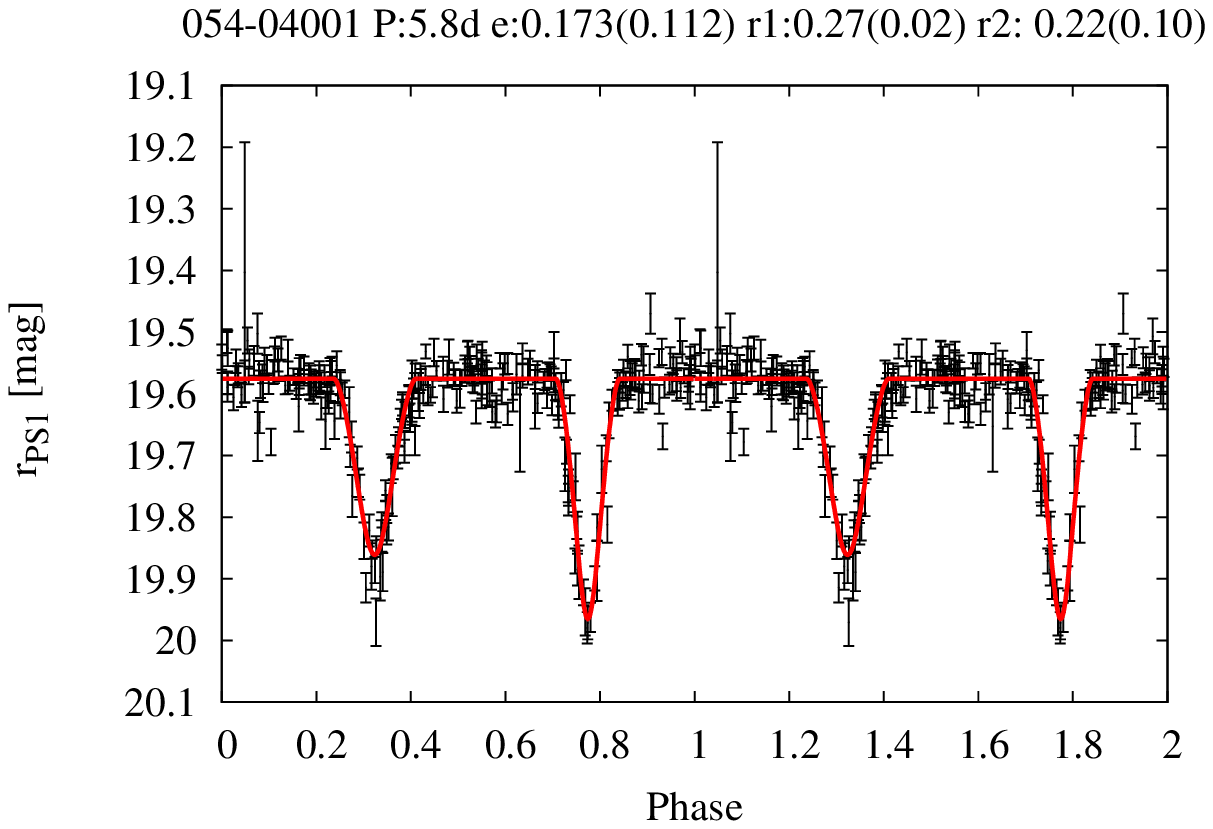}
    \includegraphics[scale=0.4]{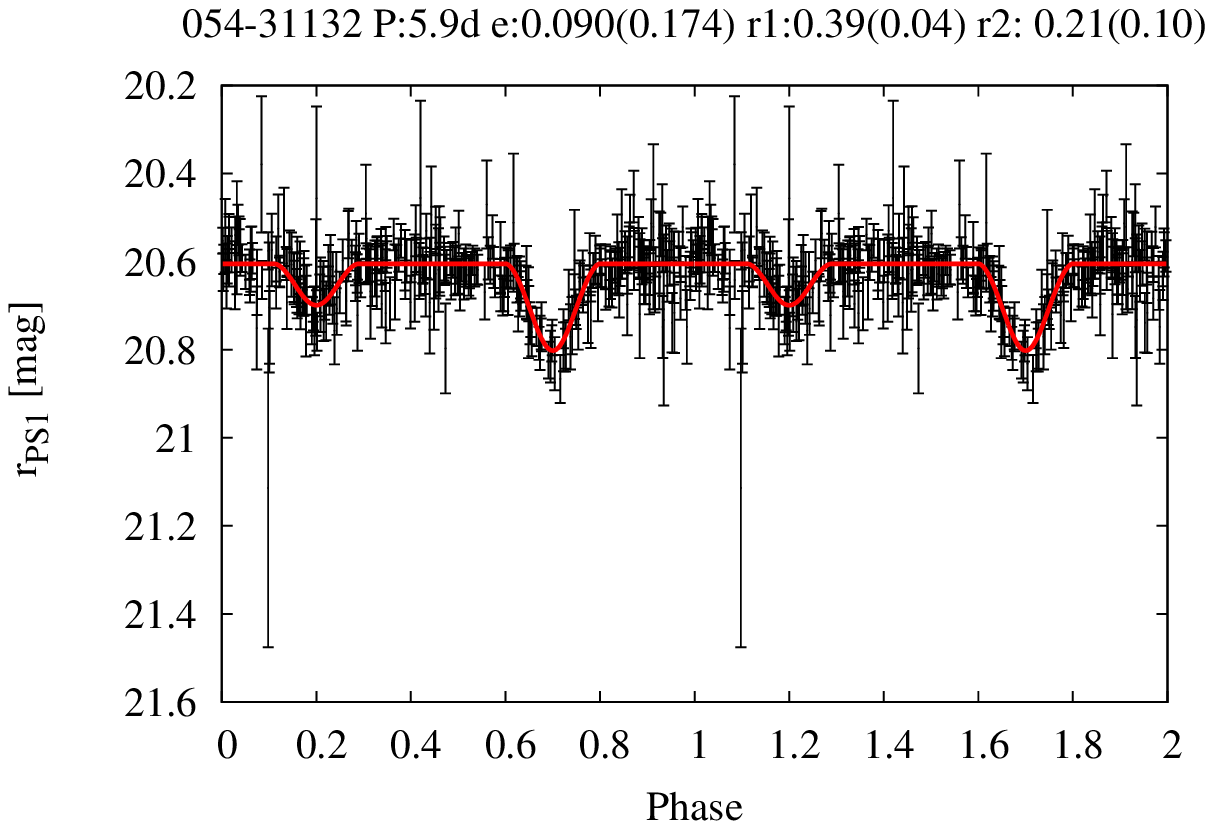}
    \includegraphics[scale=0.4]{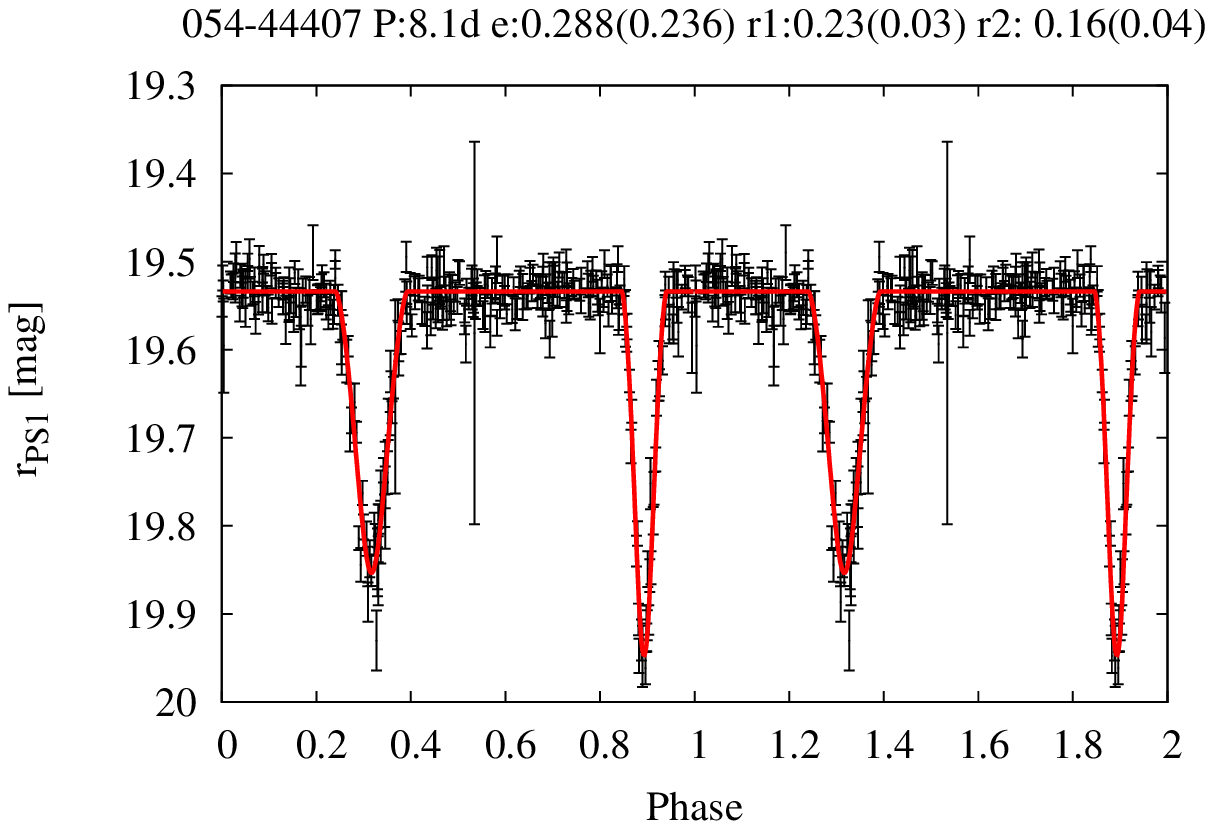}
    \includegraphics[scale=0.4]{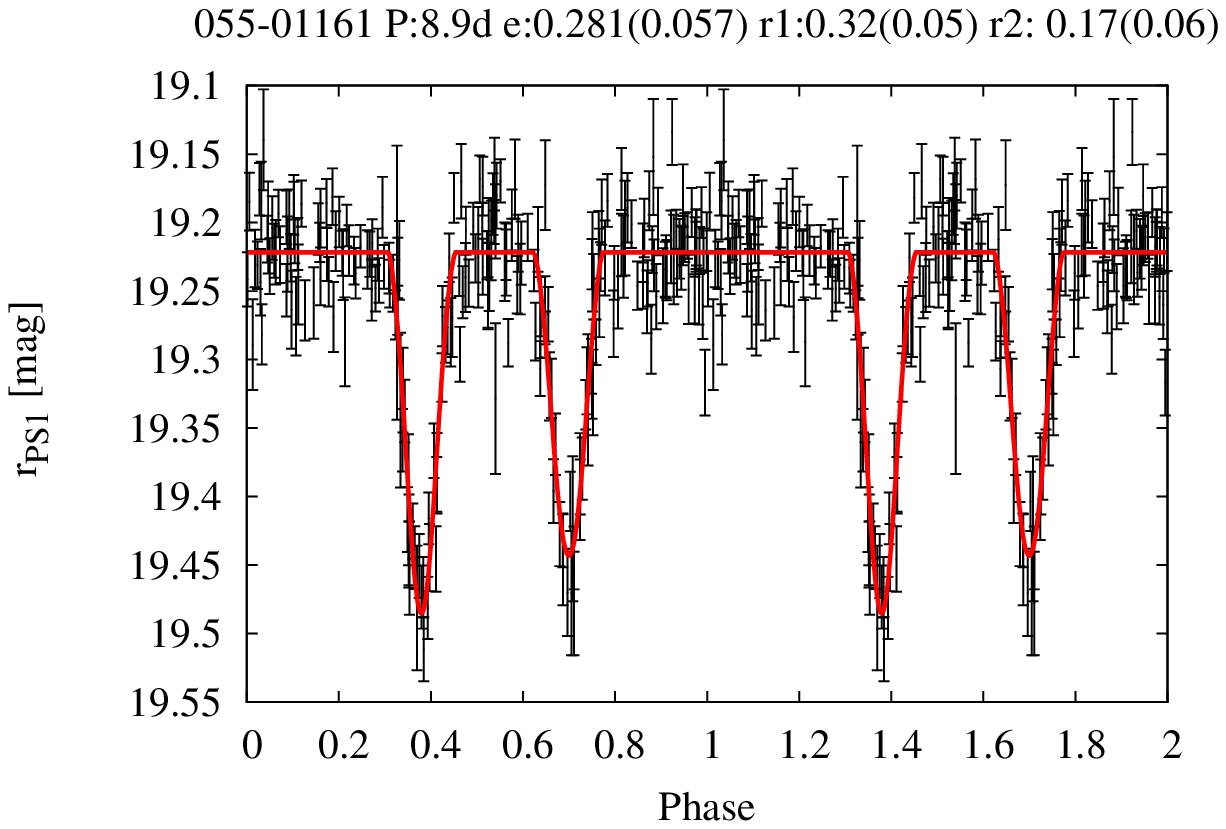}
    \includegraphics[scale=0.4]{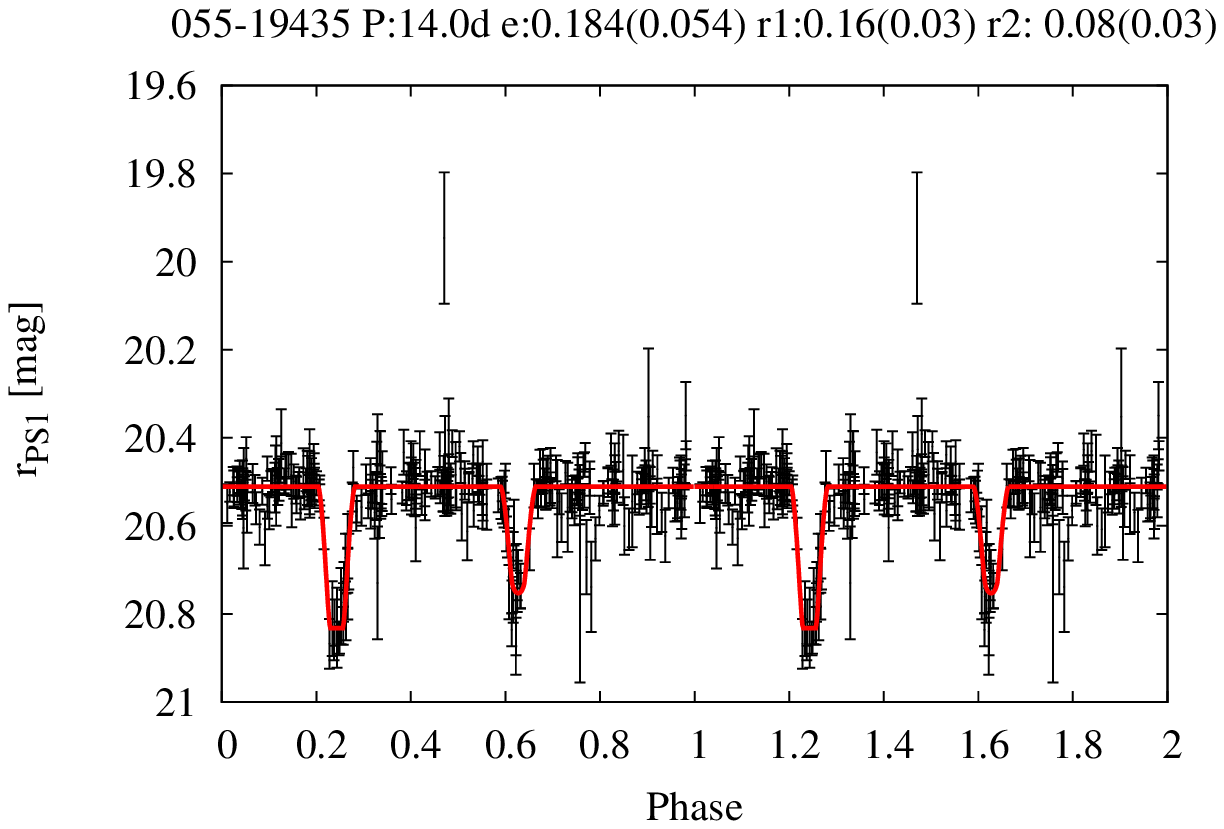}
    \includegraphics[scale=0.4]{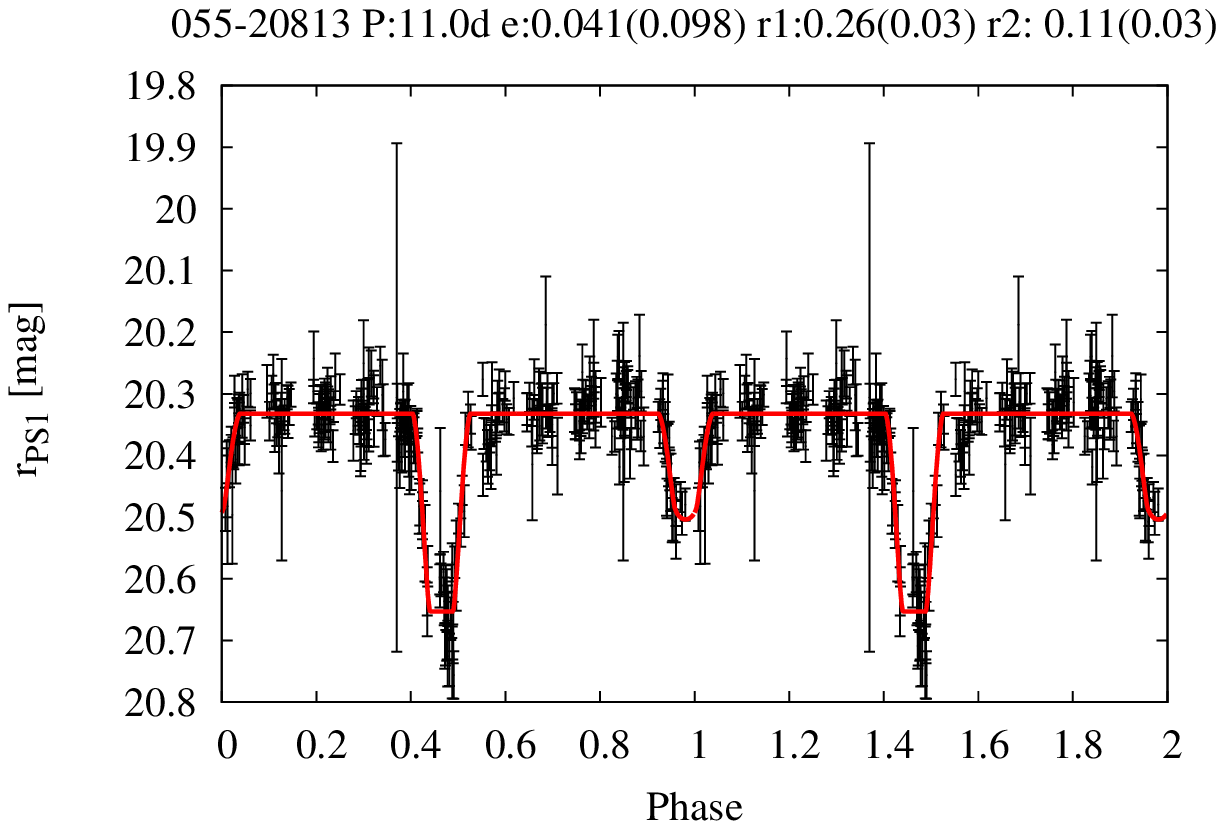}
    \includegraphics[scale=0.4]{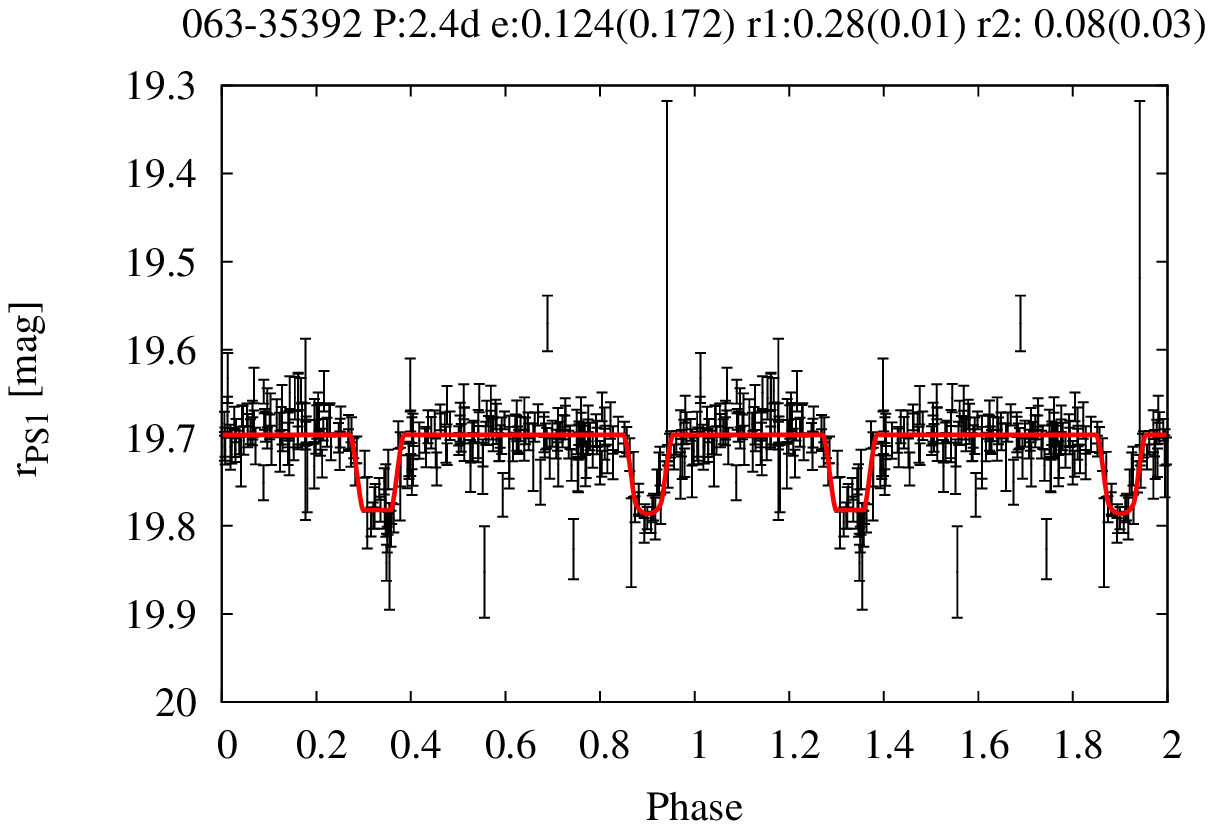}
    \caption{Lightcurves and best-fitted DEBiL model for the 13 detached binaries. The parameters in top of each panel are explained in Table \ref{tab.deb}, see also text.}
    \label{fig.deb}  
\end{figure*}

We cross-match our detached binaries with the catalog of Local Group Survey 
\citep{2006AJ....131.2478M}, and among the 298 binaries in our sample 
we find 13 candidates brighter than 20.5 mag in V, 
for which we could obtain radial velocity information from 8-10 m class telescopes
in a reasonable amount of time. We further fit their light curves with the 
Detached Eclipsing Binary Light curve fitter (DEBiL) developed by \cite{2005ApJ...628..411D}. 
Provided with an eclipsing binary light curve and its period, DEBiL can determine the 
eccentricity ($e$), the radius of the primary and secondary component in units of 
semi-major axis ($R_1$/$a$ and $R_2$/$a$), and the inclination angle (sin$i$) of 
the binary system in a robust manner. The results of the best-fitted model 
are shown in Table \ref{tab.deb}. Please note that the eccentricity is always 
fitted, but we only show the ones with errors smaller than the value of $e$. 
For the cases where the errors are larger than $e$, we consider the eccentricity 
as insignificant and show the best-fit value plus the error as an upper limit in Table \ref{tab.deb}. 
Note that in such cases, it can be either that there is negligible eccentricity, 
or that the semi-major axis is aligned to our line-of-sight, hence the eccentricity effect on
the light curve is compromised. DEBiL can only give symmetric errors, which are shown in Table 
\ref{tab.deb} following the best-fitted values of each parameter.
The light curves, as well as the best-fitted DEBiL 
models, are shown in Fig. \ref{fig.deb}

\begin{figure*}[!h]
    \centering
    \includegraphics[scale=0.7]{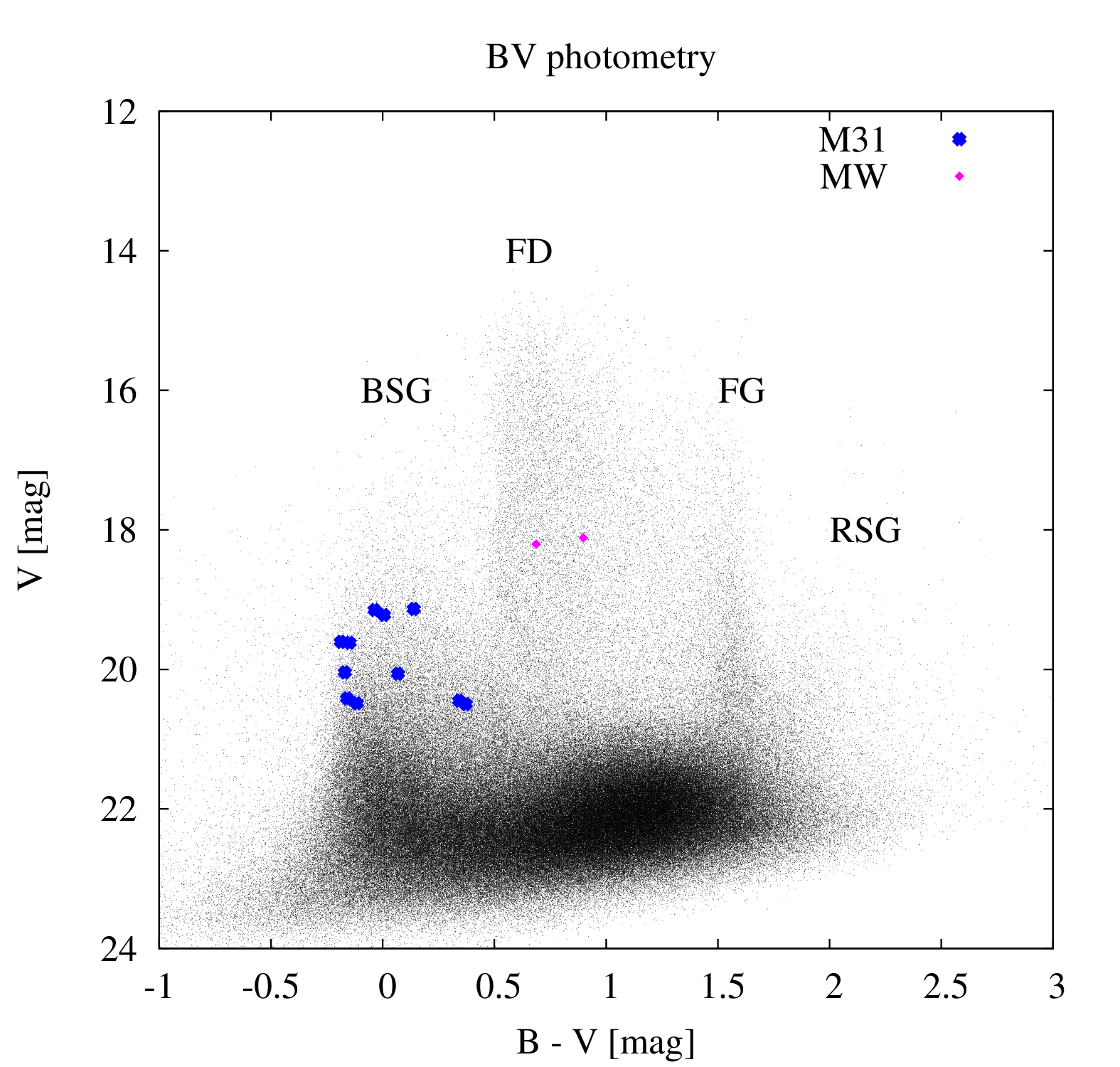}
    \caption{Color magnitude diagram for the 13 bright detached binaries. The underlaying black 
dots are photometric measurements taken from the Local Group Galaxy Survey \citep{2006AJ....131.2478M}.
Different M31 and Milky Way populations are clearly separated, e.g. blue super giants (BSG) and 
red super giants (RSG) in M31, foreground dwarves (FD) and foreground giants (FG) from Milky Way.
The two brightest detached binaries (marked in pink rhombi) from our sample exhibit relative red color, fit into the 
foreground dwarf scenario, while the remaining 11 systems (marked in blue dots) are bluer, and better fit into the 
M31 blue giants scenario.}
    \label{fig.cmd}
\end{figure*}

Among the 13 bright detached binary systems, we find two of them are too bright to 
fit into M31 scenario, hence could be foreground stars. To assess their origin as Milky Way foreground
stars, we investigate the color magnitude diagram using B-V v.s. V, where the M31 blue super giants, 
the foreground dwarves and giants from Milky Way are clearly separated, as shown in \cite{2006AJ....131.2478M}. 
We find that the two brightest systems, 034-19151 and 036-15741 exhibit relatively red color and fit
into the Milky Way foreground dwarf scenario, while the remaining 11 systems are bluer and better fit
into the M31 blue giants scenario, as shown in Fig. \ref{fig.cmd}.

\begin{sidewaystable*}[!h]
\caption{Best-fit DEBiL model of bright detached binary systems.}
\begin{tabular}{c|rcrcccccr}
\hline
Name      & RA(J2000) & Dec(J2000) & Period & $e$  & $R_1$/$a$       & $R_2$/$a$        & sin($i$)  & V      & B-V       \\
          & [deg]     & [deg]      & [day]  &      &                 &                 &            & [mag]  & [mag]     \\
\hline
026-01331 &  9.8894899 & 40.3386213 & 13.7 & 0.27$\pm$0.09 & 0.16$\pm$0.03 & 0.15$\pm$0.03 & 0.99$\pm$0.01 & 19.151 & -0.036 \\ 
026-08840 & 10.1470368 & 40.4485614 & 1.7 & 0.09$\pm$0.02 & 0.42$\pm$0.12 & 0.14$\pm$0.03 & 0.99$\pm$0.34 & 20.484 & -0.116 \\ 
026-31174 & 10.1376832 & 40.6107178 & 4.4 & 0.10$\pm$0.09 & 0.28$\pm$0.03 & 0.16$\pm$0.03 & 1.00$\pm$0.02 & 20.042 & -0.169 \\ 
027-24278 &  9.7924434 & 40.6230922 & 6.8 & $\le$0.44 & 0.28$\pm$0.01 & 0.06$\pm$0.04 & 0.99$\pm$0.07     & 20.061 &  0.068 \\  
034-19151 & 11.2190806 & 41.0353092 & 1.1 & $\le$0.19 & 0.19$\pm$0.01 & 0.11$\pm$0.05 & 1.00$\pm$0.01     & 18.113 &  0.898 \\
036-15741 & 10.1831493 & 40.8530845 & 0.7 & $\le$0.07 & 0.31$\pm$0.02 & 0.26$\pm$0.05 & 0.99$\pm$0.01     & 18.204 &  0.687 \\
054-04001 & 11.0970184 & 41.4524026 & 5.8 & 0.17$\pm$0.11 & 0.27$\pm$0.02 & 0.22$\pm$0.10 & 0.98$\pm$0.01 & 19.221 &  0.006 \\ 
054-31132 & 11.0230636 & 41.6583295 & 5.9 & $\le$0.26 & 0.39$\pm$0.04 & 0.21$\pm$0.10 & 0.95$\pm$0.05     & 20.417 & -0.159 \\  
054-44407 & 11.3321220 & 41.7514398 & 8.1 & 0.29$\pm$0.24 & 0.23$\pm$0.03 & 0.16$\pm$0.04 & 0.99$\pm$0.01 & 19.605 & -0.187 \\ 
055-01161 & 10.5500807 & 41.4274429 & 8.9 & 0.28$\pm$0.06 & 0.32$\pm$0.05 & 0.17$\pm$0.06 & 0.98$\pm$0.03 & 19.134 &  0.139 \\ 
055-19435 & 10.7472466 & 41.6185303 & 14.0 & 0.18$\pm$0.05 & 0.16$\pm$0.03 & 0.08$\pm$0.03 & 1.00$\pm$0.01 & 20.450 &  0.342 \\ 
055-20813 & 10.7648537 & 41.6311487 & 11.0 & $\le$0.14 & 0.26$\pm$0.03 & 0.11$\pm$0.03 & 1.00$\pm$0.02     & 20.498 &  0.372 \\ 
063-35392 & 11.5440668 & 42.2123537 & 2.4 & $\le$0.29 & 0.28$\pm$0.01 & 0.08$\pm$0.03 & 1.00$\pm$0.05     & 19.618 & -0.148 \\
\hline
\hline
\end{tabular}
\label{tab.deb}
\end{sidewaystable*}

\clearpage

\section{Summary}
\label{sec.sum}
We have established a customized pipeline that generates difference imaging light-curves for the PS1 M31 data. Possible variable candidates are selected with a simple $\chi^2$ criterion. We apply a modified box-fitting algorithm to determine their period and identify eclipsing binary candidates. These candidates are further classified into categories of detached, semi-detached, and contact systems with Fourier decomposition, which is carried out for M31 binaries for the first time. The period distribution shows detached and semi-detached systems 
are  longer, peaked at $>$ 10 days, while the contact systems have rather short periods. The only exception is a contact system with period larger than 10 days. Such a system can be explained by a semi-detached configuration, where one of the component is Roche-lobe filling, while the other component is considerablly smaller or cooler. The red color of this systems indeed points towards the semi-detached scheme. 

Following the classification, we cross-match our detached binaries with the Local Group Galaxy Survey, and select 13 systems brighter than 
20.5 mag in V, rendering its spectroscopic follow-up with 8-10m class ground-based telescopes feasible. We note that the distance to M31 
has been determined to $\sim$4\% accuracy using two binary systems (Vilardell et al. 2010). However, these two binaries are semi-detached,
which complicate the light curve analysis with distortions and reflection effects due to proximity of the components. In our detached 
binary sample, there are several systems exhibit eccentric orbits, where the components can be considered well separated, and reduce
the complexity of light curve modelling. 

\section{Acknowledgments}
We would like to thank the referee for the insightful comments.
This work was supported by the DFG cluster of excellence 'Origin and Structure of the Universe' (www.universe-cluster.de) and its Munich Institute for Astro and Particle Physics (MIAPP, www.munich-iapp.de). RPK acknowledges support by the National Science Foundation under grant AST-1008798. The Pan-STARRS1 Surveys (PS1) have been made possible through contributions of the Institute for Astronomy, the University of Hawaii, the Pan-STARRS Project Office, the Max-Planck Society and its participating institutes, the Max Planck Institute for Astronomy, Heidelberg and the Max Planck Institute for Extraterrestrial Physics, Garching, The Johns Hopkins University, Durham University, the University of Edinburgh, Queen's University Belfast, the Harvard-Smithsonian Center for Astrophysics, the Las Cumbres Observatory Global Telescope Network Incorporated, the National Central University of Taiwan, the Space Telescope Science Institute, the National Aeronautics and Space Administration under Grant No. NNX08AR22G issued through the Planetary Science Division of the NASA Science Mission Directorate, the National Science Foundation under Grant No. AST-1238877, the University of Maryland, and Eotvos Lorand University (ELTE).

\end{document}